\newcommand \be{\begin{equation}}
\newcommand \ba{\begin{eqnarray}}
\newcommand \ea{\end{eqnarray}}
\newcommand \ee{\end{equation}}

\documentclass[12pt,aip,
apl,
reprint
]
{revtex4-1}
\usepackage{float}
\usepackage{color}
\usepackage{epsfig}
\begin{document}

\title{Urbach tail studies by luminescence filtering in moderately doped bulk InP}
\affiliation{Department of Electrical and Computer Engineering,
State University of New York at Stony Brook, Stony Brook, NY,
11794-2350}
\author{Arsen V. Subashiev}
\email[Electronic mail: ]{subashiev@ece.sunysb.edu}
\author{Oleg Semyonov}
\author{Zhichao Chen}
\author{Serge Luryi}

\begin{abstract}
 The shape of the photoluminescence line registered from a side edge
of InP wafer is studied as function of the distance from the
excitation spot. The observed red shift in the luminescence maximum
is well described by radiation filtering and is consistent with the
absorption spectra. Our method provides an independent and accurate
determination of the Urbach tails in moderately doped
semiconductors.
\end{abstract}

\maketitle Studies of the optical absorption spectra near the
interband absorption edge are widely used for characterization of
semiconductor materials. \cite{EuSe,TlGaSe,GdH3} The spectral
dependence at low-energy absorption edge is well approximated by the
Urbach exponential decay. \cite{Urbach} However, the interband
absorption usually overlaps with residual absorption by free
carriers, masking the true dependence of the absorption tail. In
doped samples, the absorption tail is additionally broadened and
this effect is also masked by the residual electronic absorption.

Available theories\cite{Gauss,Lax,RjJohn,Greef} do not provide
description of the interband absorption tailing in the entire energy
range from interband to deep tails. Still, they give an insight on
the nature of the bandgap fluctuations causing the tailing. Tailing
with Gaussian-like asymptotics is characteristic of classical
potential fluctuations,\cite{Gauss} whereas exponential decrease
with square-root energy dependence in the exponent is indicative of
quantum effects in the band tailing.\cite{Lax,RjJohn} Temperature
variations of the tailing are accounted for by the adiabatic
potential of thermally excited phonons. \cite{Greef}

In a limited experimental range near the fundamental edge the
observed absorption spectra may not differ noticeably from the
Urbach law,\cite{RjJohn} but the temperature and the concentration
dependence of the tailing parameters can be very informative.
Therefore, accurate studies of the tailing dependence are highly
desirable.

In this letter we describe an alternative experimental method for
studying the semiconductor absorption edge by measuring the red
shift of the peak of the luminescence line, registered from the side
edge of the wafer.  This shift is sensitive to the sample
transparency at the peak wavelength, which is in the region where
residual absorption dominates.  We show that for moderately doped
$n$-InP wafers this technique provides an accurate determination of
the Urbach tailing energy.

\begin{figure}[t,b]
\epsfig{figure=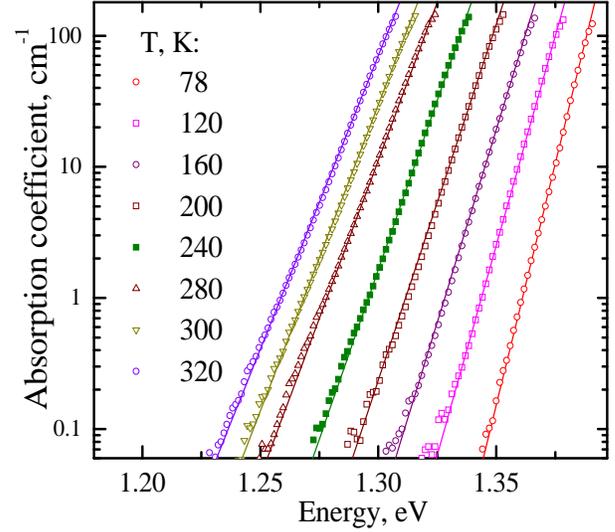,width=7.9cm,height=7.2cm} \caption[]
{(Color online) Absorption spectra of $n$-doped InP sample
($n=3\times 10^{17}$ cm$^{-3}$) at different temperatures (dots);
the lines show fits to Eq. (\ref{UrbTail}).} \label{AbsSpec}
\end{figure}
We used 350 $\mu$m-thick InP wafers, doped $n$-type (S) in the range
$n=2\times 10^{17}$ to $6\times 10^{18}$ cm$^{-3}$ and measured the
reflection and the transmission spectra to evaluate the absorption
coefficient $\alpha$.\cite{Semyon} For all moderately doped samples,
$n \le 2\times 10^{18}$ cm$^{-3}$, in the temperature range 78 to
320 K, the absorption edge exhibits an Urbach-type energy dependence
in the range $\alpha = 10$ to 100 cm$^{-1}$. For lower doped
samples, $n < 10^{18}$ cm$^{-3}$, the observed Urbach tail extends
deeper into the band gap. One can recover the interband absorption
in the bandgap by subtracting the residual (free-carrier) absorption
which is essentially constant in this energy region. The resulting
red-wing interband absorption spectra are presented in Fig.
\ref{AbsSpec} for a sample with $n=3\times 10^{17}$ cm$^{-3}$ at
several temperatures. The spectra clearly conform to the Urbach law,
\be \alpha = \alpha_0 \exp \left(- \frac{E-E_g}{\Delta(n,T)}\right),
\label{UrbTail} \ee where $\Delta(n,T)$ is the Urbach tail parameter
and $E_g$ is the bandgap energy. Matching the values of $E_g(T)$,
well-known\cite{Vurgaft} for undoped InP, gives  $\alpha_0=1.1\times
10^4$ cm$^{-1}$; this value gives a good fit in a wide temperature
range $T = 0$ to 1000 K.\cite{JohnsTielde} The physical
interpretation of $\alpha_0$ as the value of $\alpha$ at $E=E_g$
(i.e. above the steep slope) suggests that it should not vary with
the concentration at a moderate doping level, when the Fermi level
is still below the conduction band edge. The experimentally observed
temperature variations of $E_g(T)$ and $\Delta(n,T)$, are
proportional to the population of thermally excited phonons and  can
be estimated\cite{JohnsTielde, Chung} using Einstein's model for the
phonon spectrum. Variations of $E_g$ with the doping concentration
reflect the effect of band-gap narrowing combined with Fermi energy
shift due to filling of the impurity band,\cite{Bugaj} while the
increase of $\Delta$ is attributed to the combined effect of the
adiabatic random phonon potential and the random potential produced
by concentration fluctuations.\cite{Greef}

\begin{figure*}[t]
\epsfig{figure=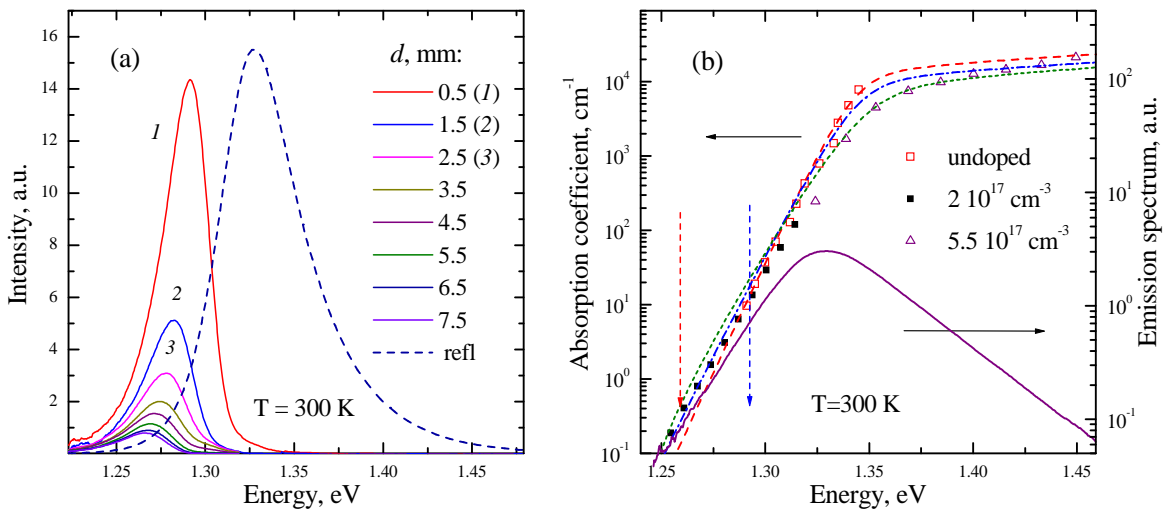,width=15.5cm,height=6.8cm} \caption[]
{(Color online) (a) Luminescence  spectra for $n$-InP sample,
$n=3\times 10^{17}$ cm$^{-3}$ at $T=300$ K at different increasing
distances $d$ between the excitation spot and the wafer edge; the
dashed line shows the emission spectrum measured in the reflection
geometry\cite{Semyon}; (b) Reflection luminescence spectrum on a
logarithmic scale together with the absorption spectra for several
doping levels. Dashed arrows indicate the energy range of the
variation in the luminescence spectral maximum} \label{SpecvsX}
\end{figure*}
The luminescence spectra were registered from the side edge of the
wafer for varying distances $d$ between the excitation spot and the
edge. The laser excitation energy was chosen close to the absorption
edge to ensure large enough excitation volume.\cite{Semyon}
Luminescence spectra for a sample with $n=3\times 10^{17}$ cm$^{-3}$
are presented in Fig. \ref{SpecvsX} (a) for several values of $d$.
Also shown (by a dashed line) is the luminescence emission spectrum,
as observed in the reflection geometry.\cite{Semyon} The red shift
of the emission line maximum with increasing $d$ is clearly seen.

Figure \ref{SpecvsX} (b) shows the absorption spectra\cite{Semyon}
for several doping levels in a broad energy range that includes the
Urbach tail region. Two vertical dashed arrows indicate the range of
the luminescence peak positions shown in Fig. \ref{SpecvsX} (a).
Also depicted is the reflection luminescence spectrum on a
logarithmic scale ($n=3\times 10^{17}$ cm$^{-3}$, $T=300$ K). It
exhibits the exponential decay both in the red and the blue wings,
and is well described by the van Roosbroek-Shockley
quasi-equilibrium relation, \cite{VRSh} viz.  $ S_0 (E) =
\alpha(E)E^2\exp\left[- E/(kT_{\rm eff})\right]$. Here $T_{\rm eff}$
is an effective temperature that can be estimated from the
exponential slope at the blue wing of the spectrum.

\begin{figure}[b]
\epsfig{figure=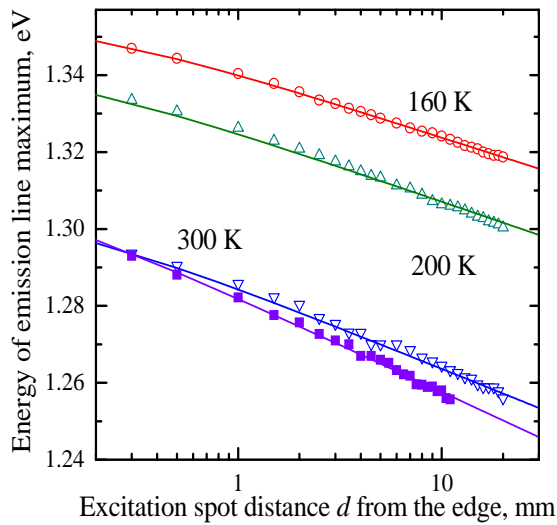,width=7.4cm,height=7.1cm} \caption[]
{(Color online)  Experimentally observed shift of the luminescence
peak with the distance $d$ from the sample edge for two samples with
$n=3 \times 10^{17}$ cm$^{-3}$ (open dots) and $n= 2\times 10^{18}$
cm$^{-3}$ (full dots); the lines show the dependence given by Eq.
(\ref{parEa}) with parameters listed in Table \ref{table1}.}
\label{MaxSpec}
\end{figure}

Figure \ref{MaxSpec} shows the dependence of the luminescence peak
position $E_{\rm max}$ on the distance $d$ between the excitation
spot and the wafer edge. The observed dependences $E_{\rm max}(d)$
for several temperatures fit accurately to an expression of the form
\be E_{\rm max}(d) = E_g-\Delta \ln [ \alpha_0 (d+d_{\rm min})/a],
\label{parEa} \ee where $\Delta$ is the Urbach tail parameter and
$d_{\rm min}$ is a small fitting parameter, $d_{\rm min} \le 0.17$
mm for all samples. The latter reflects details of the experimental
geometry (finite width and depth of the excitation spot) and is of
no importance for distances $d$ in the range of 1 to 20 mm, i.e. for
$d \gg d_{\rm min} $. Taking the values of $E_g$ from the observed
dependence $\alpha (E)$, cf. Eq. (\ref{UrbTail}), we find that the
only remaining parameter is $a$. This parameter depends on the
temperature and doping. Table \ref{table1} lists the obtained values
of $a$  for the sample with $n=3\times 10^{17}$ cm$^{-3}$ at several
temperatures and also for $n=2\times 10^{18}$ cm$^{-3}$ at $T=300$
K.
\begin{table}
\caption{\label{table1}Parameters of the luminescence spectra and
the relation $E_{\rm max}(d)$, Eqs. (\ref{parEa}) and (\ref{maxF}).}
\begin{ruledtabular}
\begin{tabular}{lllllll}
$T, K$&$E_g$, eV& $\Delta$, meV &$\Delta'$, meV&$a$&$\Delta/\Delta'$\\
\hline
160\footnotemark[1]& 1.397& 7.4 &13 & 0.58&0.57\\
200\footnotemark[1]&1.386 & 7.9 & 14& 0.54&0.56\\
300\footnotemark[1]&1.355 &9.4 & 15 &  0.63&0.63\\
300\footnotemark[2]&1.361&10.6&16 &  0.68&0.67
\end{tabular}
\end{ruledtabular}
\footnotetext[1]{$n=3\times 10^{17}$  cm$^{-3}$}
\footnotetext[2]{$n=2\times 10^{18}$  cm$^{-3}$}
\end{table}

Next we show that the observed $E_{\rm max}(d)$, including the
values of $a$, can be reproduced in a simple model that attributes
the luminescent peak shift to wavelength-dependent filtering of
outgoing radiation by the sample absorption. We assume that the
position of the peak observed at distance $d$ from the excitation
spot is determined by the transparency $T(E,d)$ of InP wafer to the
luminescence spectrum $S_0(E)$. In other words, the observed
spectrum near its maximum should be described by the product
$S(E,d)=S_0(E)\times T(E,d)$. The strong refraction of outgoing
radiation and a relatively small observation angle ensure a small
and constant range of the angles of incidence. Therefore the $d$
dependence of the transparency $T(E,d)$ reflects one-dimensional
attenuation of light \be T(E,d)= \exp\left[-\alpha(E)d\right].
\label{transp} \ee The outgoing spectra are not influenced by the
surface reflection, but may be modified by multiple interband
re-absorption of photons which leads to the creation of new minority
carriers and new radiative emission events (photon recycling).
However, if the distance $d$ is much larger than the width of the
minority carrier distribution in the source (including the
broadening of this distribution by photon-assisted diffusion), then
the spectrum near its peak is quite insensitive to these processes.
The maximum of the transmitted spectrum at varying $d$ can be found
from $d S(E,d)/dE=0$. Using (\ref{transp}), this equation can be
rewritten in the form
\be \frac {d \ln\left[S_0(E)\right]}{dE}|_{\rm max} = d \times \frac
{d \alpha(E)}{dE}|_{\rm max}. \label{eqmax} \ee
To evaluate the left-hand side of Eq. (\ref{eqmax}),  we use the
experimental exponential dependence of the luminescence spectra in
the red wing. These spectra may be modulated by some factor
reflecting the wavelength dependence of the radiation escape
probability, but this does not affect the red-wing exponential
decay.  The intrinsic emission spectrum should be closely similar to
the luminescence spectrum measured with high-energy excitation in
the reflection geometry, where the observed spectrum is not
influenced by the diffusion and filtering effects.

The decay in the red wing below the band edge is of the form \be
S_0(E) = S_{0,g} \exp \left(- \frac{E-E_g}{\Delta'}\right),
\label{expdep} \ee where $\Delta'$ is another Urbach-like tailing
parameter. The experimentally observed values of $\Delta'$ are
listed in Table \ref{table1}. Using Eqs. (\ref{UrbTail},
\ref{expdep}) to calculate the derivatives in Eq. (\ref{eqmax}), we
find \be E_{\rm max}(d) = E_g -\Delta \ln\left[\frac{\Delta'}
{\Delta}(\alpha_0 d) \right] \label{maxF}\ee Comparing  Eq.
(\ref{maxF}) with the empirical dependence (\ref{parEa}) gives a
physical interpretation to parameter $a$, viz. $a=\Delta/ \Delta'$.
As seen from the last two columns of Table \ref{table1}, this
interpretation has excellent agreement with experiment for both
samples and all temperatures.

For all studied cases, the values of  $\Delta$ obtained from the
slope of $\alpha$ and the slope of the dependence of $E_{\rm max}$
on $\ln(d)$ are very close, the difference never exceeding 0.2 meV.
Thus, the described luminescence method provides an independent way
of measuring the tailing parameters. This method can be
indispensable (in fact, the only available) in the case when the
residual absorption is strong.

In low-doped crystals the Urbach tail is known to be due to the
electron-phonon interaction, which implies a certain temperature
dependence\cite{Urbach,JohnsTielde} of the tailing parameter
$\Delta$. The doping effects in moderately- and highly-doped
crystals are more complicated. While $\Delta$ grows with
doping,\cite{Greef} the absorption spectrum is at the same time
blue-shifted by the Fermi energy of the majority carriers (the
Moss-Burstein shift). The smearing of absorption is then modified by
the temperature spread of the majority carrier energy distribution.
E.g., in doped GaAs samples, the absorption edge is often described
by non-Urbach exponential tails of Gaussian type or Halperin-Lax
type.\cite{Gauss,Lax} These are difficult to distinguish with
traditional absorption studies because of the residual absorption.
Our evaluation from the luminescence experiments is more accurate in
the region of small $\alpha$ -- allowing to study and identify
different types of tailing.

Finally, we note that in moderately doped III-V semiconductors the
high quantum radiative efficiency results in high photon recycling
\cite{Semyon,Lush2} that gives rise to a photon-enhanced minority
carrier transport and can broaden the initial hole distribution over
an enlarged diffusion length.\cite{Dumke,vonRoss} More accurate
consideration\cite{Serge10} shows that the recycling-induced carrier
transport should be viewed as anomalous diffusion -- due to the
extremely long photon propagation in the transparency region at the
red wing of the emission spectrum. In our experiments, this
phenomenon manifests itself as strongly enhanced values of $a$ in
low-doped samples at $T=78$ K. The approach described in the present
work is well suited to study these anomalous diffusion effects, as
will be reported separately.

This work was supported by the Domestic Nuclear Detection Office of
the Department of Homeland Security, by the Defense Threat Reduction
Agency through its basic research program, and by the New York State
Office of Science, Technology and Academic Research through the
Center for Advanced Sensor Technology at Stony Brook.

\end{document}